# Service-Oriented Fast Frequency Response from Flexible Loads and Energy Storage in Low-Inertia Power Systems

Xiaojie Tao[a,*], Rajit Gadh[a]

[a] Smart Grid Energy Research Center, Mechanical and Aerospace Engineering Department,

University of California, Los Angeles, CA 90095, USA

*Corresponding Author: Xiaojie Tao, University of California, Los Angeles, CA 90095, USA

Email: xiaojietao@g.ucla.edu, taoxiaojie04@gmail.com

**Abstract**: The increasing penetration of inverter-based renewable generation has significantly reduced system inertia, making modern power grids more vulnerable to rapid frequency deviations following disturbances. While a wide range of flexible resources—including electric vehicles (EVs), data centers, and battery energy storage systems (BESS)—have demonstrated the physical capability to provide fast frequency response (FFR), existing studies primarily focus on individual resource performance or controller-level designs. A systematic framework that translates heterogeneous FFR capabilities into deployable, system-level frequency services remains largely unexplored.

This paper proposes a service-oriented coordination framework for fast frequency response from flexible loads and energy storage, bridging the gap between physical capability assessment and grid-operational utilization. The framework decomposes frequency support into multiple time-critical service layers based on response speed, power capacity, and energy sustainability, and dynamically allocates FFR responsibilities among heterogeneous resources accordingly. By explicitly accounting for response latency, saturation limits, and energy constraints, the proposed approach enables coordinated dispatch that prioritizes ultra-fast resources for initial frequency arrest while leveraging slower but energy-rich resources to sustain recovery.

# 1. Introduction

The rapid growth of inverter-based renewable energy sources has fundamentally altered the dynamic behavior of modern power systems [1]-[10]. As conventional synchronous generation is displaced by wind and solar resources, overall system inertia continues to decline, leading to faster and deeper frequency deviations following disturbances [11]-[15]. Under such low-inertia conditions, traditional governor-based primary frequency control—typically responding on the order of several seconds—may no longer be sufficient to arrest frequency drops within secure limits [16]-[18]. Consequently, fast frequency response (FFR), characterized by sub-second to few-second active power adjustments triggered by local frequency measurements, has become a critical requirement for maintaining frequency stability in future power grids [19]-[23].

In parallel with these challenges, a diverse set of flexible resources has emerged across the electricity system [24]-[26]. Electric vehicles (EVs) equipped with smart charging or vehicle-to-grid capability, data centers with controllable IT workloads and uninterruptible power supply (UPS) systems, and battery energy storage systems (BESS) all possess the ability to adjust active power at timescales relevant to FFR [27]-[30]. Numerous studies have demonstrated that these resources can individually provide rapid frequency support and improve key performance metrics such as frequency nadir and rate-of-change-of-frequency (RoCoF) [31]-[35]. However, most existing work focuses on evaluating the physical capability or control design of individual resource types, often under fixed participation assumptions or resource-specific operating scenarios.

While such capability-based assessments are essential, they do not fully address a critical system-level question: how should heterogeneous fast-responding resources be deployed and coordinated to provide reliable and efficient frequency support under practical grid operation? In real power systems, frequency control is not delivered by isolated devices, but by structured services with implicit requirements on response speed, duration, reliability, and substitutability [38]. Treating all fast-responding resources as equivalent contributors to a single, undifferentiated FFR action neglects fundamental differences in response latency, power saturation, and energy sustainability, and may result in inefficient utilization of scarce ultra-fast resources or unnecessary reliance on high-cost assets [39].

Recent research has begun to explore coordinated frequency response from multiple resource types, demonstrating that aggregation can outperform single-resource strategies. Nevertheless, coordination is often implemented through fixed droop gains, static priority rules, or centralized optimization formulations that remain closely tied to specific resource models [40]. As a result, the connection between physical response capability and deployable frequency services remains implicit, limiting the applicability of these approaches to real-world system operation, aggregation, and market design [41].

This paper addresses this gap by proposing a service-oriented coordination framework for fast frequency response from flexible loads and energy storage. Instead of treating FFR solely as a control action, the proposed framework explicitly structures frequency support into multiple time-critical service layers based on response speed, power capability, and energy duration. Heterogeneous resources are dynamically coordinated to provide complementary contributions, with ultra-fast devices prioritizing initial frequency arrest and slower but energy-rich resources sustaining the response and facilitating recovery. By accounting for practical constraints such as response latency, saturation limits, and energy availability, the framework bridges the gap between physical capability assessment and system-level deployment of frequency services.

The effectiveness of the proposed framework is evaluated using a modified IEEE 39-bus test system under low-inertia operating conditions. Through comparative case studies and sensitivity analyses, the paper quantifies improvements in frequency stability as well as the marginal value and substitutability of different resource types under varying availability scenarios. The results demonstrate that structured, service-based coordination can achieve superior frequency performance while reducing dependence on scarce ultra-fast resources, offering important insights for system operators, aggregators, and future fast frequency response mechanisms.

The remainder of this paper is organized as follows. Section 2 introduces the requirements of fast frequency response from a service perspective and outlines the proposed service decomposition. Section 3 presents the coordination framework and resource allocation logic. Section 4 describes the system model and case study setup. Section 5 discusses simulation results and sensitivity analyses, and Section 6 concludes the paper with key findings and implications for future grid operation.

## 2. Fast Frequency Response as a Grid Service

### 2.1 Limitations of Capability-Based FFR Modeling

Fast frequency response (FFR) is commonly modeled as a frequency-dependent active power adjustment delivered by fast-acting resources following a disturbance. In most existing studies, FFR capability is characterized by parameters such as droop gain, response delay, and power limits, and system-level performance is evaluated by aggregating these responses in a unified control loop. While this approach provides valuable insights into the physical capability of individual devices, it implicitly assumes that all fast-responding resources contribute equivalently to frequency control.

In practice, however, heterogeneous FFR-capable resources differ fundamentally in their dynamic characteristics. Power-electronic energy storage systems can respond within tens of milliseconds but are often constrained by limited energy capacity. Flexible loads such as data centers and electric vehicle fleets typically exhibit longer response delays but can sustain power modulation over extended durations. Aggregating such resources under a single, undifferentiated FFR signal may lead to inefficient utilization, premature saturation of ultra-fast assets, or suboptimal frequency recovery behavior.

Moreover, grid operators do not deploy frequency response solely based on physical capability, but rather through implicitly defined services with specific performance expectations. These expectations include not only response magnitude, but also activation speed, duration, reliability, and interaction with other control layers. A purely capability-based formulation therefore lacks an explicit mapping between device-level response and system-level service provision.

**2.2 Service-Oriented View of Fast Frequency Response**

To address these limitations, this paper adopts a service-oriented perspective on fast frequency response. From a system operation standpoint, FFR can be interpreted as a set of structured frequency services delivered over distinct time windows following a disturbance.

Each service layer is characterized by its response speed requirement, expected contribution duration, and functional role in stabilizing system frequency.

Under this perspective, the objective of coordination is not to maximize instantaneous power injection from all available resources, but to ensure that frequency stability requirements are met with appropriate allocation of response responsibilities across time. This distinction is particularly important in low-inertia systems, where the initial seconds following a disturbance are critical, but sustained support is equally necessary to prevent secondary frequency dips or excessive reliance on conventional reserves.

By explicitly structuring FFR into service layers, heterogeneous resources can be deployed according to their comparative advantages rather than treated as interchangeable actuators. This approach enables systematic coordination, improves transparency in resource utilization, and provides a natural interface between physical response capability and deployable grid services.

**2.3 Decomposition of Fast Frequency Response Services**

Based on typical frequency dynamics following a generation loss, FFR is decomposed in this work into three conceptual service layers, as illustrated in Fig. 1.

**(1) Ultra-fast frequency arrest service.**

This service operates within the first few hundred milliseconds after a disturbance and is primarily responsible for reducing the initial rate of change of frequency (RoCoF) and arresting rapid frequency decline. Resources participating in this layer must exhibit minimal response latency and high power ramping capability. Converter-interfaced battery energy storage systems and data center UPS inverters are well suited for this role, while energy limitations restrict the duration of sustained response.

**(2) Fast sustained support service.**

Following the initial arrest, frequency stabilization requires continued active power support over a time horizon of several seconds. This service layer bridges the gap between ultra-fast response and slower primary control actions. Resources such as data centers with controllable IT workloads and aggregated EV fleets can contribute effectively in this window, offering moderate response speed combined with greater energy sustainability.

**(3) Energy-following frequency support.**

Beyond the immediate stabilization phase, frequency recovery benefits from energy-rich resources capable of sustaining power modulation without rapid saturation. Although this layer overlaps temporally with traditional primary frequency control, flexible loads and distributed energy resources can reduce the burden on conventional generation by providing supplementary support during recovery.

These service layers are not mutually exclusive and may overlap in time. However, their explicit definition clarifies the functional role of different resource types and enables structured coordination across heterogeneous assets.

**2.4 Implications for Resource Coordination**

The proposed service decomposition has important implications for coordinating flexible loads and energy storage. First, it highlights that faster response does not necessarily imply greater overall value; ultra-fast resources are most valuable during the initial disturbance window but should be preserved for critical events. Second, slower resources with larger energy capacity can provide substantial system value when deployed appropriately to sustain response and facilitate recovery.

By aligning resource participation with service-layer requirements, coordination strategies can prioritize speed where necessary and energy where available, avoiding unnecessary overreaction and improving overall efficiency. This service-oriented formulation also enables systematic evaluation of resource substitutability, allowing system operators to assess how reductions in one resource class may be compensated by increased participation from others.

In the following section, this service-oriented view is translated into a coordination framework that dynamically allocates fast frequency response responsibilities among heterogeneous resources based on their real-time availability and dynamic characteristics.

## 3. Service-Oriented Coordination Framework

### 3.1 Framework Overview

Building on the service-oriented interpretation of fast frequency response introduced in Section 2, this section presents a coordination framework that translates heterogeneous resource capabilities into structured frequency services. The objective of the framework is not to enforce a unified control law across all resources, but to dynamically allocate fast frequency response responsibilities in a manner consistent with service-layer requirements and resource-specific characteristics.

The proposed framework adopts a hierarchical yet lightweight structure. System frequency deviation is measured locally and used as a common triggering signal, while coordination decisions are based on predefined service roles rather than device-level optimization. This design choice avoids excessive communication and computational complexity, making the framework suitable for real-time deployment by system operators or aggregators.

At a high level, the framework consists of three functional components:

1. **Service-layer definition**, which specifies response requirements across different time windows;
2. **Resource capability mapping**, which characterizes each resource's suitability for each service layer;
3. **Dynamic allocation logic**, which assigns response responsibility based on availability, speed, and energy constraints.

This structure explicitly separates *what services are needed* from *how individual devices deliver them*, enabling scalable coordination across diverse flexible resources.

**3.2 Resource Capability Representation**

Each fast-responding resource is characterized using a compact set of parameters that capture its relevance to fast frequency response services. Rather than relying on detailed device models, the framework represents each resource $i$ using the following attributes:

- **Response latency** $\tau_i$, representing the delay between frequency deviation detection and effective power response;
- **Maximum deliverable power** $P_i^{\max}$, defining instantaneous response capability;
- **Sustainable energy budget** $E_i$, reflecting how long the response can be maintained without saturation;
- **Operational availability** $A_i$, accounting for practical constraints such as state of charge, workload flexibility, or connection status.

These attributes jointly define a *capability envelope* for each resource. Fast-acting energy storage systems typically exhibit small $\tau_i$ and large $P_i^{\max}$ but limited $E_i$, whereas flexible loads such as data centers and EV fleets exhibit larger $\tau_i$ but significantly greater energy sustainability. This abstraction allows heterogeneous devices to be compared and coordinated on a common basis without enforcing uniform control structures.

**3.3 Service-Based Allocation Logic**

The coordination logic assigns resources to service layers based on their relative capability envelopes and real-time availability. Instead of allocating response solely based on instantaneous frequency deviation, the framework distributes responsibility across time according to service priorities.

During the **ultra-fast frequency arrest phase**, resources with minimal latency are prioritized. Allocation is constrained to a limited subset of ultra-fast devices to ensure rapid RoCoF mitigation while avoiding unnecessary energy depletion. As the system transitions into the **fast sustained support phase**, participation from slower but more energy-rich resources is gradually increased, allowing ultra-fast assets to reduce output or recover energy reserves. Finally, during the **energy-following support phase**, sustained contributors maintain frequency stabilization while conventional primary control actions engage.

This staged allocation approach ensures that response speed and energy sustainability are both respected. Importantly, allocation decisions are adaptive: if a particular resource class becomes unavailable or constrained, responsibility can be redistributed across remaining resources within the same service layer or shifted to adjacent layers.

**3.4 Handling Saturation and Resource Substitutability**

A key advantage of the service-oriented framework is its ability to manage resource saturation and substitutability in a transparent manner. When a resource approaches its power or energy limits, its participation weight within the corresponding service layer is reduced, triggering compensatory contributions from other eligible resources.

This mechanism enables systematic evaluation of trade-offs between different resource classes. For example, reduced availability of ultra-fast storage can be partially offset by increased participation from flexible loads during later response stages, albeit with potential impacts on frequency nadir. Conversely, increased deployment of fast energy storage can reduce the required contribution from slower resources, preserving flexibility for other grid services.

By explicitly structuring these interactions, the framework provides operators with actionable insight into how resource portfolios influence frequency stability, beyond simple aggregate capacity metrics.

**3.5 Implementation Considerations**

The proposed coordination framework is designed to complement existing frequency control mechanisms rather than replace them. Service-layer allocation operates on top of local droop or frequency-based controllers, preserving device autonomy and minimizing implementation barriers. Communication requirements are limited to availability signaling and coarse coordination commands, avoiding reliance on centralized optimization or high-bandwidth data exchange.

From an operational perspective, the framework can be implemented by system operators, aggregators, or large flexible-load coordinators as an intermediate control layer.

Its service-oriented formulation also aligns naturally with emerging discussions on differentiated fast frequency response products, providing a conceptual foundation for future grid codes and procurement mechanisms.

The following section applies this framework to a representative low-inertia power system to evaluate its effectiveness and to quantify the system-level benefits of service-based coordination.

## 4. Case Study

**4.1 Test System Description**

The proposed service-oriented coordination framework is evaluated using a modified IEEE 39-bus New England test system, which is widely adopted for frequency stability studies. The system consists of ten synchronous generators, 39 buses, and 19 aggregated loads, and provides a realistic representation of transmission-level dynamics under large disturbance events.

To emulate low-inertia operating conditions characteristic of renewable-dominated power systems, the total system inertia is uniformly reduced across synchronous generators while maintaining the original network topology and steady-state power flow. This approach preserves the structural characteristics of the test system while allowing controlled assessment of frequency dynamics under reduced inertia scenarios. All synchronous generators are equipped with standard primary frequency control via turbine governors, representing baseline system response.

System frequency is evaluated using the center-of-inertia (COI) formulation, which reflects the aggregate dynamic behavior of the interconnected system and avoids sensitivity to local oscillations.

**4.2 Integration of Flexible Frequency Response Resources**

Three classes of fast frequency response resources are integrated into the test system to reflect realistic heterogeneity:

1. **Battery Energy Storage Systems (BESS)**, representing ultra-fast, power-dense resources with limited energy capacity;
2. **Data centers**, equipped with UPS inverters and controllable IT workloads, providing fast but energy-moderate response;
3. **Electric vehicle (EV) fleets**, aggregated through smart charging or vehicle-to-grid control, offering slower but energy-rich flexibility.

Each resource class is connected to a representative load bus selected based on electrical strength and load concentration, ensuring that frequency measurements are consistent with system-wide dynamics. Local frequency deviation is used as the triggering signal for all resources, avoiding reliance on centralized frequency measurement or high-bandwidth communication.

Rather than enforcing uniform dynamic models, each resource follows its native response characteristics—such as inverter dynamics for BESS and UPS systems, or delayed aggregate response for EV fleets—while participating in the coordination framework through service-layer allocation signals.

**4.3 Disturbance Scenarios and Evaluation Metrics**

To assess framework performance, a representative generation loss event is simulated by tripping the largest synchronous generator in the system. This disturbance produces a sudden power imbalance and rapid frequency decline, challenging the system's ability to maintain frequency security under low inertia.

The following performance metrics are used for evaluation:

- **Frequency nadir**, indicating the depth of frequency decline;
- **Rate-of-change-of-frequency (RoCoF)**, reflecting the severity of the initial disturbance;
- **Frequency recovery time**, defined as the time required to return within a specified deadband around nominal frequency;
- **Resource utilization profiles**, capturing power output and saturation behavior across service layers.

These metrics collectively capture both short-term and sustained frequency stability performance, aligning with the service-oriented objectives introduced in Sections 2 and 3.

**4.4 Case Definitions**

Four cases are defined to isolate the impact of service-oriented coordination:

- **Case 1: Baseline (No FFR)**

  No fast frequency response is provided by flexible resources. System frequency relies solely on inertia and conventional primary control.

- **Case 2: Unstructured Aggregation**

  All flexible resources respond simultaneously using fixed droop-based control without service-layer differentiation.

- **Case 3: Partial Service Allocation**

  Resources are assigned to service layers based on response speed, but without dynamic adjustment for saturation or availability.

- **Case 4: Proposed Service-Oriented Coordination**

  The full framework is implemented, dynamically allocating response responsibility across service layers based on latency, power capability, and energy availability.

This progression allows direct comparison between capability-based aggregation and structured, service-oriented coordination.

**4.5 Parameter Selection and Implementation Notes**

Resource parameters are selected to reflect representative deployment scenarios reported in the literature, including response delays, power limits, and energy constraints for each resource class. To ensure fair comparison across cases, total installed flexible capacity is held constant, and only coordination logic is varied.

The service-layer coordination operates as a supervisory mechanism, adjusting participation weights while preserving local frequency-based controllers. Anti-saturation logic is enforced to prevent excessive depletion of limited-energy resources, and gradual handoff between service layers is implemented to ensure smooth transitions.

All simulations are conducted with a sufficiently small integration time step to capture sub-second dynamics, and each case is simulated over an identical time horizon to enable consistent comparison.

**4.6 Summary**

This case study framework is designed to evaluate not only whether fast frequency response resources improve system performance, but how different coordination philosophies influence frequency stability, resource utilization, and substitutability. By focusing on service-level deployment rather than device-level tuning, the case study directly reflects the objectives of the proposed framework and sets the stage for the comparative results presented in the following section.

## 5. Simulation Results and Discussion

**5.1 System Frequency Response Under Different Coordination Strategies**

Figure 2 compares the system frequency trajectories following a large generation loss under the four cases defined in Section 4. In the baseline case without fast frequency response, the reduced system inertia leads to a rapid frequency decline, resulting in a deep frequency nadir and elevated RoCoF. Conventional primary control alone is insufficient to arrest the initial frequency drop within secure limits.

When all flexible resources respond simultaneously using unstructured aggregation (Case 2), frequency performance improves compared with the baseline. However, the initial frequency nadir remains relatively deep, and pronounced oscillations appear during the recovery phase. This behavior reflects premature saturation of ultra-fast resources and inadequate sustained support once fast-acting devices reach their limits.

The partial service allocation case (Case 3) further improves frequency nadir by prioritizing faster resources during the initial response window. Nevertheless, because participation weights are fixed, resource saturation still occurs during prolonged support, resulting in suboptimal recovery behavior and inefficient utilization of energy-rich resources.

In contrast, the proposed service-oriented coordination framework (Case 4) yields the best overall frequency performance. The frequency nadir is significantly improved, RoCoF is reduced, and recovery to nominal frequency is smoother and faster. These improvements are achieved not by increasing total response capacity, but by aligning resource participation with service-layer requirements across time.

**5.2 Role of Service Layers in Frequency Stabilization**

To better understand the observed improvements, Figure 3 illustrates the power contributions of different resource classes under the proposed framework. During the ultra-fast frequency arrest phase, battery energy storage systems and data center UPS units dominate the response, delivering rapid power injection to suppress RoCoF. Their contribution is intentionally constrained in duration to preserve energy reserves.

As the system transitions into the fast sustained support phase, participation from data centers and EV fleets increases, while ultra-fast resources gradually reduce output. This staged handoff prevents abrupt power withdrawal and avoids secondary frequency dips, a common issue observed in unstructured aggregation.

During the energy-following phase, energy-rich resources maintain support until conventional primary control stabilizes the system. This layered behavior confirms that frequency stability is achieved not through maximal instantaneous response, but through temporally structured deployment of heterogeneous resources.

**5.3 Resource Utilization and Saturation Behavior**

Figure 4 compares resource utilization profiles across coordination strategies. In unstructured aggregation, ultra-fast storage devices rapidly reach power and energy limits, leading to early saturation and limited contribution beyond the initial response window. This behavior necessitates greater reliance on conventional generation during recovery.

Under service-oriented coordination, saturation is actively managed by adjusting participation weights as resources approach operational limits. Ultra-fast devices are relieved once their primary service role is fulfilled, while slower resources absorb greater responsibility. This dynamic allocation reduces stress on any single resource class and improves overall system robustness.

These results demonstrate that effective coordination can extend the useful contribution of limited-energy assets without compromising frequency security.

**5.4 Substitutability and Marginal Value of Resource Classes**

A key advantage of the proposed framework is its ability to quantify the substitutability of different fast frequency response resources. Sensitivity analyses are conducted by systematically reducing the availability of one resource class while maintaining the same disturbance conditions.

Results show that reductions in ultra-fast storage capacity lead to measurable degradation in RoCoF and frequency nadir. However, increased participation from data centers and EV fleets partially compensates for this loss during later response stages, mitigating overall performance degradation. Conversely, reductions in energy-rich flexible loads primarily affect recovery time rather than initial frequency arrest.

These findings indicate that different resource classes provide complementary, rather than interchangeable, value. Ultra-fast resources are critical for initial stabilization, while energy-rich resources are essential for sustained support. The service-oriented framework makes these trade-offs explicit, enabling informed decisions regarding resource procurement and deployment.

**5.5 Discussion and Practical Implications**

The results highlight several important implications for low-inertia grid operation. First, simply increasing the number of fast-responding devices does not guarantee improved frequency stability; coordination structure plays a decisive role. Second, service-oriented deployment allows system operators to extract greater value from existing resources, potentially reducing the need for costly ultra-fast storage investments.

From an operational perspective, the framework provides a transparent mechanism for mapping heterogeneous device capabilities to differentiated frequency services. This perspective aligns naturally with emerging discussions on fast frequency response products and may inform future grid codes, aggregator designs, and procurement strategies.

Overall, the results confirm that frequency stability in low-inertia systems depends not only on the availability of flexible resources, but on their structured and coordinated deployment as time-critical grid services.

## 6. Conclusion and Future Work

This paper proposed a service-oriented coordination framework for fast frequency response (FFR) from flexible loads and energy storage in low-inertia power systems. Moving beyond conventional capability-based formulations, the framework explicitly structures frequency support into time-critical service layers and coordinates heterogeneous resources according to their response speed, power capability, and energy sustainability.

This perspective bridges the gap between device-level response capability and deployable, system-level frequency services.

Through case studies on a modified IEEE 39-bus system, the proposed framework demonstrated clear advantages over unstructured aggregation and fixed-allocation strategies. By aligning resource participation with service-layer requirements, the framework achieved improved frequency nadir, reduced rate-of-change-of-frequency, and smoother recovery behavior without increasing total flexible capacity. The results highlight that effective frequency stability depends not only on the availability of fast-responding resources, but also on their structured deployment across time.

The analysis further revealed the complementary roles of different resource classes. Ultra-fast resources such as battery energy storage systems and data center UPS units are critical for initial frequency arrest, while energy-rich flexible loads—including data centers and electric vehicle fleets—play a dominant role in sustaining frequency support and facilitating recovery. By making these roles explicit, the service-oriented formulation enables systematic assessment of resource substitutability and marginal value, providing actionable insights for system operators and planners.

From an operational standpoint, the proposed framework offers a transparent and scalable approach to coordinating heterogeneous flexible resources without relying on centralized optimization or detailed device-level modeling. Its hierarchical and service-based structure aligns naturally with emerging discussions on differentiated fast frequency response products and may inform future grid codes, aggregation strategies, and procurement mechanisms in low-inertia grids.

Future work will extend the proposed framework in several directions. First, integration with market-based procurement and pricing mechanisms will be explored to translate service-layer contributions into economically efficient incentives. Second, interactions between fast frequency response services and other grid services—such as primary frequency control, reserve provision, and congestion management—will be investigated. Finally, validation using large-scale system models and field data will be pursued to assess scalability and robustness under realistic operating conditions.

**CRediT authorship contribution statement**

**Xiaojie Tao**: Conceptualization, Methodology, Software, Validation, Formal analysis, Investigation, Writing - Original Draft, Writing - Review & Editing, Visualization. **Rajit Gadh**: Supervision, Project administration, Funding acquisition, Investigation, Resources, Writing – Review & Editing.

**Declaration of competing interest**

The authors declare that they have no known competing financial interests or personal relationships that could have appeared to influence the work reported in this paper.

**Data Availability**

Data will be made available on request.

**Acknowledgements**

This work was supported by the Department of Mechanical and Aerospace Engineering, University of California, Los Angeles, under grant numbers 69763, 77739, and 45779.

**References**

[1] Thiruvengadam P, Thiruvengadam A, Ryskamp R, Besch M, Perhinschi M, Demirgök B, et al. A Novel Approach to Test Cycle-Based Engine Calibration Technique Using Genetic Algorithms to Meet Future Emissions Standards. SAE International Journal of Engines 2020;13. https://doi.org/10.4271/03-13-04-0036.

[2] Sabzehgar R, Roshan YM, Fajri P. Modeling and Control of a Multifunctional Three-Phase Converter for Bidirectional Power Flow in Plug-In Electric Vehicles. Energies 2020;13:2591. https://doi.org/10.3390/en13102591.

[3] Decentralized Failure-Tolerant Optimization of Electric Vehicle Charging | IEEE Journals & Magazine | IEEE Xplore n.d. https://ieeexplore.ieee.org/document/9431209 (accessed November 8, 2025).

[4] Xu S, Li J, Zhang X, Zhu D. Research on Optimal Driving Torque Control Strategy for Multi-Axle Distributed Electric Drive Heavy-Duty Vehicles. Sustainability 2024;16:7231. https://doi.org/10.3390/su16167231.

[5] Almeida L, Soares A, Moura P. A Systematic Review of Optimization Approaches for the Integration of Electric Vehicles in Public Buildings. Energies 2023;16:5030. https://doi.org/10.3390/en16135030.

[6] Wang Y, Sheikh O, Hu B, Chu C-C, Gadh R. Integration of V2H/V2G Hybrid System for Demand Response in Distribution Network. City of Los Angeles Department; 2014. https://doi.org/10.1109/SmartGridComm.2014.7007748.

[7] Hashemi S, Arias NB, Andersen PB, Christensen B, Traholt C. Frequency Regulation Provision Using Cross-Brand Bidirectional V2G-Enabled Electric Vehicles. 2018 IEEE International Conference on Smart Energy Grid Engineering (SEGE) 2018. https://doi.org/10.1109/sege.2018.8499485.

[8] A Hardware-in-the-Loop Test on the Multi-Objective Ancillary Service by In-Vehicle Batteries: Primary Frequency Control and Distribution Voltage Support | IEEE Journals & Magazine | IEEE Xplore n.d. https://ieeexplore.ieee.org/document/8892569?denied= (accessed November 8, 2025).

[9] Quasi Single-Stage Three-Phase Filterless Converter for EV Charging Applications | IEEE Journals & Magazine | IEEE Xplore n.d. https://ieeexplore.ieee.org/document/9648000?denied= (accessed November 8, 2025).

[10] Mak H-Y, Tang R. Collaborative Vehicle-to-Grid Operations in Frequency Regulation Markets. M&SOM 2024;26:814–33. https://doi.org/10.1287/msom.2022.0133.

[11] Divani MY, Najafi M, Ghaedi A, Gorginpour H. Security-constrained optimal scheduling and operation of island microgrids considering demand response and electric vehicles. International Transactions on Electrical Energy Systems 2021;31:e13178. https://doi.org/10.1002/2050-7038.13178.

[12] Narasimhulu N, Awasthy M, Pérez de Prado R, Divakarachari PB, Himabindu N. Analysis and Impacts of Grid Integrated Photo-Voltaic and Electric Vehicle on Power Quality Issues. Energies 2023;16:714. https://doi.org/10.3390/en16020714.

[13] DeForest N, MacDonald JS, Black DR. Day ahead optimization of an electric vehicle fleet providing ancillary services in the Los Angeles Air Force Base vehicle-to-

grid demonstration. Applied Energy 2018;210:987–1001. https://doi.org/10.1016/j.apenergy.2017.07.069.

[14] Ahmed M, Abouelseoud Y, Abbasy NH, Kamel SH. Hierarchical Distributed Framework for Optimal Dynamic Load Management of Electric Vehicles With Vehicle-to-Grid Technology. IEEE Access 2021;9:164643–58. https://doi.org/10.1109/ACCESS.2021.3134868.

[15] Loutan C, Klauer P, Chowdhury S, Hall SG, Morjaria M, Chadliev V, et al. Demonstration of Essential Reliability Services by a 300-MW Solar Photovoltaic Power Plant. SciteAi 2017. https://doi.org/10.2172/1349211.

[16] Makarov YV, Lu S, Ma J, Nguyen T. Assessing the Value of Regulation Resources Based on Their Time Response Characteristics. SciteAi 2008. https://doi.org/10.2172/946001.

[17] Scarabaggio P, Carli R, Cavone G, Dotoli M. Smart Control Strategies for Primary Frequency Regulation through Electric Vehicles: A Battery Degradation Perspective. Energies 2020;13. https://doi.org/10.3390/en13174586.

[18] Alhelou HH, Siano P, Tipaldi M, Iervolino R, Mahfoud F. Primary Frequency Response Improvement in Interconnected Power Systems Using Electric Vehicle Virtual Power Plants. World Electric Vehicle Journal 2020;11:40. https://doi.org/10.3390/wevj11020040.

[19] Arias NB, Hashemi S, Andersen PB, Traholt C, Romero R. V2G enabled EVs providing frequency containment reserves: Field results. 2018 IEEE International Conference on Industrial Technology (ICIT), Lyon: IEEE; 2018, p. 1814–9. https://doi.org/10.1109/ICIT.2018.8352459.

[20] Chen R, Bauchy M, Wang W, Sun Y, Tao X, Marian J. Using graph neural network and symbolic regression to model disordered systems. Sci Rep 2025;15:22122. https://doi.org/10.1038/s41598-025-05205-8.

[21] Mohammadi F, Nazri G-A, Saif M. A Bidirectional Power Charging Control Strategy for Plug-in Hybrid Electric Vehicles. Sustainability 2019;11:4317. https://doi.org/10.3390/su11164317.

[22] Liang H, Lee Z, Li G. A Calculation Model of Charge and Discharge Capacity of Electric Vehicle Cluster Based on Trip Chain. IEEE Access 2020;8:142026–42. https://doi.org/10.1109/ACCESS.2020.3014160.

[23] Hossain S, Rokonuzzaman M, Rahman KS, Habib AKMA, Tan W-S, Mahmud M, et al. Grid-Vehicle-Grid (G2V2G) Efficient Power Transmission: An Overview of Concept, Operations, Benefits, Concerns, and Future Challenges. Sustainability 2023;15:5782. https://doi.org/10.3390/su15075782.

[24] Dean MD, Kockelman KM. Assessing Public Opinions of and Interest in Bidirectional Electric Vehicle Charging Technologies: A U.S. Perspective. Transportation Research Record: Journal of the Transportation Research Board 2024;2678. https://doi.org/10.1177/03611981241253608.

[25] A Survey Data Approach for Determining the Probability Values of Vehicle-to-Grid Service Provision n.d. https://www.mdpi.com/1996-1073/14/21/7270 (accessed November 8, 2025).

[26] Taljegard M, Walter V, Göransson L, Odenberger M, Johnsson F. Impact of electric vehicles on the cost-competitiveness of generation and storage technologies in the

electricity system. Environ Res Lett 2019;14:124087. https://doi.org/10.1088/1748-9326/ab5e6b.

[27] Tamura S. A V2G strategy to increase the cost–benefit of primary frequency regulation considering EV battery degradation. Electrical Engineering in Japan 2020;212:11–22. https://doi.org/10.1002/eej.23270.

[28] Liu H, Qi J, Wang J, Li P, Li C, Wei H. EV Dispatch Control for Supplementary Frequency Regulation Considering the Expectation of EV Owners. IEEE Transactions on Smart Grid 2018;9:3763–72. https://doi.org/10.1109/TSG.2016.2641481.

[29] Bahmani MH, Esmaeili Shayan M, Fioriti D. Assessing electric vehicles behavior in power networks: A non-stationary discrete Markov chain approach. Electric Power Systems Research 2024;229:110106. https://doi.org/10.1016/j.epsr.2023.110106.

[30] Iqbal S, Xin A, Jan MU, Salman S, Zaki A ul M, Rehman HU, et al. V2G Strategy for Primary Frequency Control of an Industrial Microgrid Considering the Charging Station Operator. Electronics 2020;9:549. https://doi.org/10.3390/electronics9040549.

[31] Zecchino A, Prostejovsky AM, Ziras C, Marinelli M. Large-scale provision of frequency control via V2G: The Bornholm power system case. Electric Power Systems Research 2019;170:25–34. https://doi.org/10.1016/j.epsr.2018.12.027.

[32] Li ZR, Tao X, Kim C-JC. Fabrication and Evaluation of Hierarchical Superhydrophobic and Salvinia Surfaces. 2023 22nd International Conference on Solid-State Sensors, Actuators and Microsystems (Transducers), 2023, p. 1348–51.

[33] Zhang J, Xu H. Online Identification of Power System Equivalent Inertia Constant. IEEE Trans Ind Electron 2017;64:8098–107. https://doi.org/10.1109/TIE.2017.2698414.

[34] Zeng F, Zhang J, Chen G, Wu Z, Huang S, Liang Y. Online Estimation of Power System Inertia Constant Under Normal Operating Conditions. IEEE Access 2020;8:101426–36. https://doi.org/10.1109/ACCESS.2020.2997728.

[35] Izadkhast S, Garcia-Gonzalez P, Frias P. An Aggregate Model of Plug-In Electric Vehicles for Primary Frequency Control. IEEE Trans Power Syst 2015;30:1475–82. https://doi.org/10.1109/TPWRS.2014.2337373.

[36] Decentralized Vehicle-to-Grid Control for Primary Frequency Regulation Considering Charging Demands | IEEE Journals & Magazine | IEEE Xplore n.d. https://doi.org/10.1109/TPWRS.2013.2252029.

[37] Xiaojie Tao; Yaoyu Fan; Zhaoyi Ye; Rajit Gadh. Heavy-Duty Electric Vehicles Contribution for Frequency Response in Power Systems with V2G, IEEE International Conference on Green Energy and Smart Systems 2025; Long Beach, CA, USA.

[38] Tao, Xiaojie; Fan, Yaoyu; Ye, Zhaoyi; Gadh, Rajit. Heavy-Duty Electric Vehicles Contribution for Frequency Response in Power Systems with V2G. arXiv, 2025, arXiv:2512.12872

[39] Tao, Xiaojie; Gadh, Rajit. Fast Frequency Response Potential of Data Centers through Workload Modulation and UPS Coordination. arXiv, 2025, arXiv:2512.14128

[40] Tao, Xiaojie; Gadh, Rajit. Coordinated Fast Frequency Response from Electric Vehicles, Data Centers, and Battery Energy Storage Systems. arXiv, 2025, arXiv:2512.14136

[41]     Tao, Xiaojie; Fan, Yaoyu; Ye, Zhaoyi; Gadh, Rajit. Assessing the Frequency Response Potential of Heavy-Duty Electric Vehicles with Vehicle-to-Grid Integration in the California Power System. Preprint, 2025. (DOI to be added upon publication).